\documentclass{llncs}

\usepackage{cite}
\usepackage{xspace}
\usepackage{hyperref}
\usepackage{xcolor}
\usepackage{graphicx}

\newcommand{\jkind}{{\sc JKind}\xspace}
\newcommand{\kind}{{\sc Kind}\xspace}
\newcommand{\pkind}{{\sc PKind}\xspace}

\newcommand{\lustre}{{\sc Lustre}\xspace}
\newcommand{\spear}{{\sc SpeAR}\xspace}
\newcommand{\simpal}{{\sc SIMPAL}\xspace}
\newcommand{\limp}{{\sc Limp}\xspace}
\newcommand{\agree}{{\sc AGREE}\xspace}

\newcommand{\nuxmv}{{\sc nuXmv}\xspace}
\newcommand{\zustre}{{\sc Zustre}\xspace}
\newcommand{\yices}{{\sc Yices}\xspace}
\newcommand{\zthree}{{\sc Z3}\xspace}

\renewcommand{\paragraph}[1]{\vspace{5pt}\noindent {\bf #1}}

\title{The \jkind Model Checker}
\author{
  Andrew Gacek\inst{1} \and
  John Backes\inst{1} \and
  Mike Whalen\inst{2} \and
  Lucas Wagner\inst{1} \and
  Elaheh Ghassabani\inst{2}
}
\institute{
  Rockwell Collins \\
  \email{andrew.gacek@gmail.com}, \email{john.backes@gmail.com}, \email{lucas.wagner@rockwellcollins.com}
  \and
  University of Minnesota \\
  \email{ghass013@umn.edu}, \email{mwwhalen@umn.edu}
}

\begin{document}
\maketitle

\begin{abstract}
\jkind is an open-source industrial model checker developed by
Rockwell Collins and the University of Minnesota.  \jkind uses
multiple parallel engines to prove or falsify safety properties of
infinite state models. It is portable, easy to install, performance
competitive with other state-of-the-art model checkers, and has
features designed to improve the results presented to users: {\em
  inductive validity cores} for proofs and {\em counterexample
  smoothing} for test-case generation.  It serves as the back-end for
various industrial applications.
\end{abstract}

\section{Introduction}

\jkind is an
open-source\footnote{\url{https://github.com/agacek/jkind}} industrial
infinite-state inductive model checker for safety properties. Models
and properties in \jkind are specified in
\lustre~\cite{halbwachs1991ieee}, a synchronous data-flow language,
using the theories of linear real and integer arithmetic. \jkind uses
SMT-solvers to prove and falsify multiple properties in parallel.
A distinguishing characteristic of \jkind is its focus on the usability  of results. For a proven property, \jkind provides traceability between the property and individual model elements. For a falsified property, \jkind provides options for simplifying the
counterexample in order to highlight the root cause of the failure. In industrial applications, we have found these additional usability aspects to be at least as important as the primary results.
Another important characteristic of \jkind is that is it designed to be integrated directly into user-facing applications. Written in Java, \jkind runs on all major platforms and is easily compiled into other Java applications. \jkind bundles the Java-based {\sc SMTInterpol} solver and has no external dependencies. However, it can optionally call \zthree, \yices 1, \yices 2, {\sc CVC4}, and {\sc
  MathSAT} if they are available.

\section{Functionality and Main Features}

\begin{figure}[t]
  \begin{center}
    \includegraphics[clip,trim=140 220 80 140,scale=0.6]{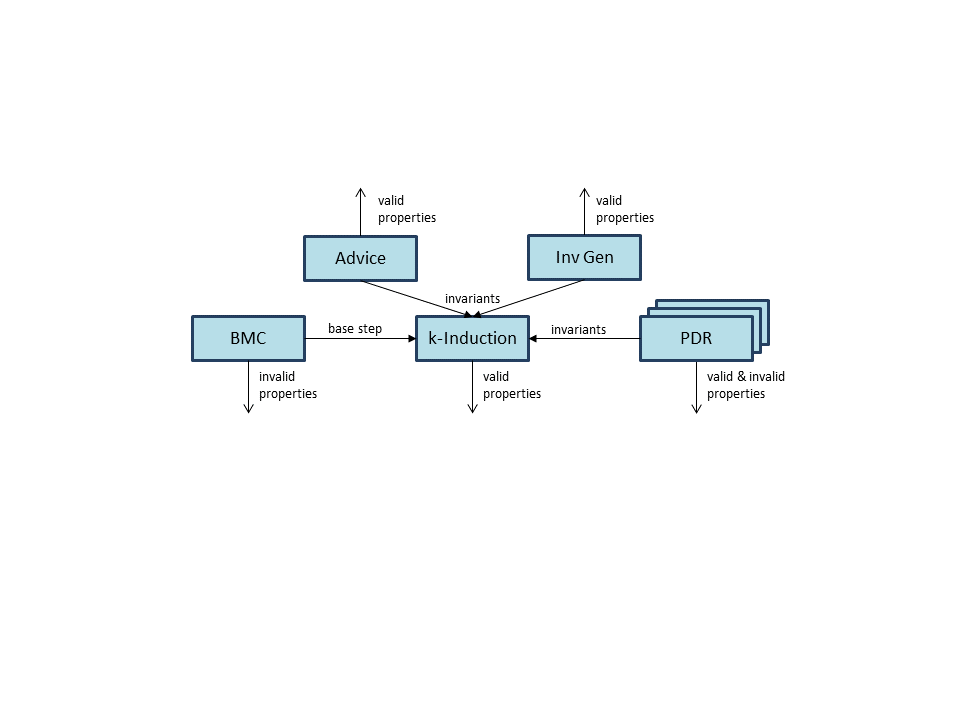}
  \end{center}
  \vspace{-2em}
  \caption{\jkind engine architecture}
  \vspace{-1em}
  \label{fig:engines}
\end{figure}

\jkind is structured as several parallel engines that coordinate to
prove properties, mimicking the design of \pkind and \kind 2~\cite{champion2016cav, kahsai2011pdmc}.
Some engines are directly responsible for proving properties, others aid that effort by generating invariants, and still others are reserved for post-processing of proof or counterexample results. Each engine can be enabled or disabled separately based on the user's needs. The architecture of \jkind allows any engine to broadcast information to the other engines (for example, lemmas, proofs, counterexamples) allowing straightforward integration of new functionality.

The solving engines in \jkind are show in
Figure~\ref{fig:engines}. The \textbf{Bounded Model Checking (BMC)}
engine performs a standard iterative unrolling of the transition
relation to find counterexamples and to serve as the base case of
$k$-induction. The BMC engine guarantees that any counterexample it
finds is minimal in length. The \textbf{$k$-Induction} engine performs
the inductive step of $k$-induction, possibly using invariants
generated by other engines. The \textbf{Invariant Generation} engine
uses a template-based invariant generation
technique~\cite{kahsai2012nfm} using its own $k$-induction loop. The
\textbf{Property Directed Reachability (PDR)} engine performs property
directed reachability~\cite{een2011fmcad} using the implicit
abstraction technique~\cite{cimatti2014tacas}. Unlike BMC and
$k$-induction, each property is handled separately by a different PDR
sub-engine. Finally, the \textbf{Advice} engine produces invariants
based on previous runs of \jkind as described in the next section.


Invariant sharing between the solvers (shown in Figure~\ref{fig:engines}) is an important part of the architecture.  In our internal benchmarking, we have found that implicit abstraction PDR performs best when operating over a single property at a time and without use of lemmas generated by other approaches.  On the other hand, the invariants generated by PDR and template lemma generation often allow $k$-induction, which operates on all properties in parallel, to substantially reduce the verification time required for models with large numbers of properties.  

\subsection{Post Processing and Re-verification}

A significant part of the research and development effort for \jkind\ has focused on
post-processing results for presentation and repeated verification of models under development.


\paragraph{Inductive Validity Cores (IVC).} For a proven property, an inductive validity core is a subset of \lustre equations from the input model for which the property still
holds~\cite{ghassabani2016fse,Ghass17AllIVCs}.  Inductive validity cores can be used for traceability from property to model elements and determining coverage of the model by a set of properties~\cite{Ghass17Cov}.  This facility can be used to automatically generate traceability and adequacy information (such as traceability matrices~\cite{fifarek2017nfm} important to the certification of safety-critical avionics systems~\cite{DO178C}).
The IVC engine uses a heuristic algorithm to efficiently produce minimal or nearly minimal cores.   In a recent experiment over a superset of the benchmark models described in the experiment in Section~\ref{sec:experiment}, we found that our heuristic IVC computation added 31\% overhead to model checking time, and yielded cores approximately 8\% larger than the guaranteed minimal core computed by a very expensive ``brute force'' algorithm.  As a side-effect, the IVC algorithm also minimizes the set of invariants used to prove a property and emits this reduced set to other engines (notably the {\em Advice} engine, described below).


\paragraph{Smoothing.}  To aid in counterexample understanding and in
creating structural coverage tests that can be more easily explained,
\jkind provides an optional post-processing step to minimize the
number of changes to input variables---{\em smoothing} the
counterexample.  For example, applied to 129 test cases generated for
a production avionics flight control state machine, smoothing
increased runtime by 40\% and removed 4 unnecessary input changes per
test case on average.  The smoothing engine uses a {\sc MaxSat} query
over the original BMC-style unrolling of the transition relation
combined with weighted assertions that each input variable does not
change on each step. The {\sc MaxSat} query tries to satisfy all of
these weighted assertions, but will break them if needed. This has the
effect of trying to hold all inputs constant while still falsifying
the original property and only allowing inputs to change when
needed. This engine is only available with SMT-solvers that support
{\sc MaxSat} such as \yices 1 and \zthree.

\paragraph{Advice.} The advice engine saves and re-uses the invariants that were used by \jkind to prove the properties of a model.  Prior to analysis, \jkind\ performs model slicing and flattening to generate a flat transition-relation model.  Internally, invariants are stored as a set of proven formulas (in the \lustre syntax) over the variables in the flattened model.  An {\em advice} file is simply the emitted set of these invariant formulas.  When a model is loaded, the formulas are loaded into memory. Formulas that are no longer syntactically or type correct are discarded, and the remaining set of formulas are submitted as an initial set of possible invariants to be proved via $k$-induction. If they are proved, they are passed along to other engines; if falsified, they are discarded.
Names constructed between multiple runs of \jkind are stable, so if a
model is unchanged, it can be usually be re-proved quickly using the
invariants and $k$-induction.  If the model is slightly changed, it is
often the case that most of the invariants can be re-proved, leading
to reduced verification times.

If the IVC engine is also enabled, then advice emits a (close to) minimal set of lemmas used for proof; this often leads to faster re-verification (but more expensive initial verification), and can be useful for examining which of the generated lemmas are useful for proofs.

\section{Experimental Evaluation}
\label{sec:experiment}
We evaluated the performance of \jkind against \kind
2~\cite{champion2016cav}, \zustre~\cite{Zustre}, Generalized PDR in
\zthree~\cite{GPDR}, and IC3 in \nuxmv~\cite{cimatti2014tacas}. We
used the default options for each tool (using {\tt check\_invar\_ic3}
for \nuxmv).  Our benchmark suite comes from~\cite{cimatti2014tacas}
and contains 688 models over the theory of linear integer
arithmetic\footnote{\url{https://es.fbk.eu/people/griggio/papers/tacas14-ic3ia.tar.bz2}.
  Note that we removed 263 duplicate benchmarks from the original
  set.}.  All experiments were performed on a 64-bit Ubuntu
17.10 Linux machine with a 12-core Intel Xeon CPU E5-1650 v3 @
3.50GHz, with 32GB of RAM and a time limit of 60 seconds per model.

\begin{figure}[t]
  \begin{center}
    \includegraphics[width=\textwidth]{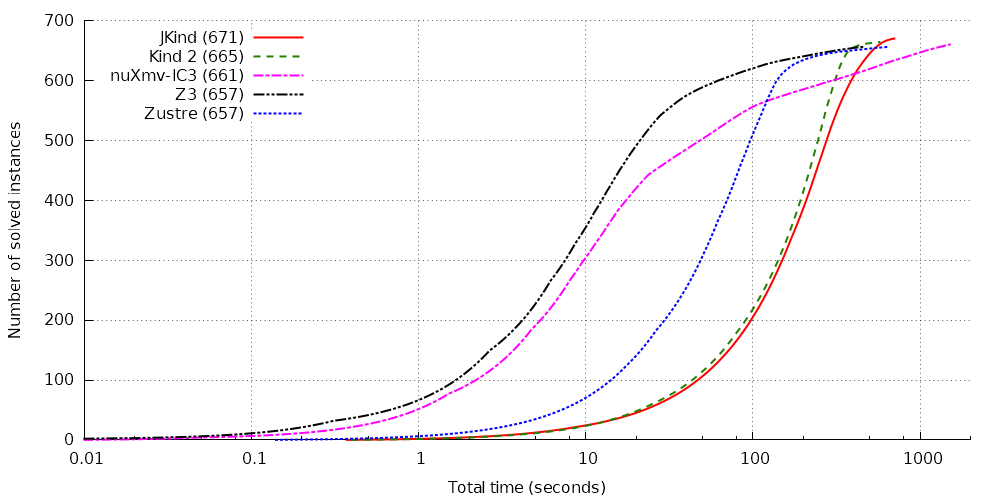}
  \end{center}
  \vspace{-2em}
  \caption{Performance benchmarks}
  \vspace{-1em}
  \label{fig:benchmark}
\end{figure}

Performance comparisons are show in
Figure~\ref{fig:benchmark}. The key describes the number of benchmarks
solved for each tool, and the graph shows the aggregate time required
for solving, ordered by time required per-problem, ordered
independently for each tool. \jkind was able to verify or falsify the
most properties, although \zthree was often the fastest tool. Many of
the benchmarks in this set are quickly evaluated: \zthree solves the
first 400 benchmarks in just over 12 seconds.  Due to \jkind's use of
Java, the JVM/\jkind startup time for an empty
model is approximately 0.35s, which leads to poor performance on small
models\footnote{Without startup time, the curve for \jkind is close to
  the curve for \zustre}.  As always, such benchmarks should be taken
with a large grain of salt.  In~\cite{champion2016cav}, a different
set of benchmarks slightly favored \kind2, and
in~\cite{cimatti2014tacas}, \nuxmv was the most capable tool.  We
believe that all the solvers are relatively competitive.

\section{Integration \& Applications}


\jkind is the back-end for a variety of user-facing applications. In this section, we briefly highlight a few of these applications and how they employ the features discussed previously.
%
%

(1) The {\em Specification and Analysis of Requirements} (\spear) tool is an open-source tool for prototyping and analysis of requirements~\cite{fifarek2017nfm}.  Starting from a set of formalized requirements, \spear uses \jkind to determine whether or not the requirements meet certain {\em properties}.  It uses IVCs to create a traceability matrix between requirements and properties, highlighting unused requirements, over-constrained properties, and other common problems. \spear also uses \jkind with smoothing for test-case generation using the Unique First Cause criteria~\cite{whalen2006issta}.
%
%

(2) The {\em Assume Guarantee Reasoning Environment} (\agree) tool is
an open-source compositional verification tool that proves properties
of hierarchically-composed models in the Architectural Analysis and
Design Language (AADL) language~\cite{cofer2012nfm,QFCS15:backes,hilt2013}.
%
%
%
\agree makes use of multiple \jkind features including smoothing to
present clear counterexamples, IVC to show requirements traceability,
and counterexample generation to check the consistency of an AADL
component's contract. \agree also uses \jkind for test-case generation
from component contracts.

(3) The {\em Static IMPerative AnaLyzer} (\simpal) tool is an
open-source tool for compositional reasoning over
software~\cite{wagner2017spin}. \simpal is based on \limp, a
\lustre-like imperative language with extensions for control flow
elements, global variables, and a syntax for specifying preconditions,
postconditions, and global variable interactions of preexisting
components. \simpal translates \limp programs to an equivalent \lustre
representation which is passed to \jkind to perform assume-guarantee
reasoning, reachability, and viability analyses.

(4) \jkind is also used by two proprietary tools used by product areas
within Rockwell Collins.  The first is a {\em Mode Transition Table}
verification tool used for the complex state machines which manage
flight modes of an aircraft.  \jkind is used to check properties and
generate tests for mode and transition coverage from \lustre models
generated from the state machines.  IVCs are used to establish
traceability, i.e. which transitions are covered by which properties.
The second is a {\em Crew Alerting System} MC/DC test-case generation
tool for a proprietary domain-specific language used for messages and
alerts to airplane pilots.  Smoothing is very important in this
context as test cases need to be run on the actual hardware where
timing is not precisely controllable. Thus, test cases with a minimum
of changes to the inputs are ideal.

\section{Related Work}
\jkind is one of a number of similar infinite-state inductive model
checkers including {\sc Kind 2}~\cite{champion2016cav},
\nuxmv~\cite{cimatti2014tacas}, \zthree with generalized
PDR~\cite{GPDR}, and \zustre~\cite{Zustre}. They operate over a
transition relation described either as a \lustre program (\kind 2,
\jkind, and \zustre), an extension of the SMV language (\nuxmv), or as
a set of Horn clauses (\zthree).  Each tool uses a portfolio-based
solver approach, with \nuxmv, \jkind, and \kind 2 all
supporting both $k$-induction and a variant of PDR/IC3.  \nuxmv also
supports guided reachability and $k$-liveness.  Other tools such as
{\sc ESBMC-DepthK} \cite{rocha2017model}, {\sc VVT}
\cite{beyer2016smt} {\sc CPAchecker}, \cite{beyer2015boosting}, {\sc
  CPROVER} \cite{brain2015safety} use similar techniques for
reasoning about C programs.

We believe that the \jkind IVC support is similar to~\emph{proof-core}
support provided by commercial hardware model checkers: Cadence Jasper
Gold and Synopsys VC Formal~\cite{hanna2015formal, jasper_gold,
  Synopsys_VC_formal}.  The proof-core provided by these tools is used
for internal coverage analysis measurements performed by the tools.
Unfortunately, the algorithms used in the commercial tool support are
undocumented and performance comparisons are prohibited by the tool
licenses, so it is not possible to compare performance on this aspect.

Previous work has been done on improving the quality of
counterexamples along various dimensions similar to the \jkind notion
of {\em smoothing}, e.g. \cite{groce2005bmc, ravi2004tacas}. Our work
is distinguished by its focus on minimizing the number of deltas in
the input values. This metric has been driven by user needs and by our
own experiences with test-case generation.

There are several tools that support reuse or exchange of verification
results, similar to our {\em advice} feature.  Recently, there has
been progress on standardized formats~\cite{Beyer:2016} of exchange
between analysis tools.  Our current advice format is optimized for
use and performance with our particular tool and designed for
re-verification rather than exchange of partial verification
information.  However, supporting a standardized format for exchanging
verification information would be a useful feature for future use.


\section{Conclusion}
\jkind is similar to a number of other solvers that each solve
infinite state sequential analysis problems. Nevertheless, it has some
important features that distinguish it. First, a focus on quality of
feedback to users for both valid properties (using IVCs) and invalid
properties (using smoothing). Second, it is supported across all major platforms and is
straightforward to port due to its implementation in Java. Third, it
is small, modular, and well-architected, allowing straightforward
extension with new engines. Fourth, it is open-source with a liberal
distribution license (BSD), so it can be adapted for various purposes,
as demonstrated by the number of tools that have incorporated it.

\paragraph{Acknowledgments}

\noindent The work presented here was sponsored by DARPA as part of the HACMS
program under contract FA8750-12-9-0179.

\bibliography{main}{}
\bibliographystyle{splncs03}

\end{document}